\documentclass[12pt]{article}
\usepackage{helvet,times,mathptm}
\usepackage{graphicx}
\begin{document}

\begin{center}
{\noindent
{\bf Inverse Vector Operators}}
\end{center}

\begin{center}
{\noindent
{\small \bf Shaon Sahoo} \footnote[1]{shaon@physics.iisc.ernet.in}}
\end{center}

\begin{center}
{\noindent 
{\small 
Department of Physics, Indian Institute of Science, Bangalore 560012, India.
\vspace{.2cm}
~\\
}}
\end{center}

{\noindent
{\bf Abstract}
{\small
In different branches of physics, we frequently deal with vector del operator 
($\vec{\nabla}$). This del operator is generally used to find curl or 
divergence of a vector function or gradient of a scalar function. In many 
important cases, we need to know the parent vector whose curl or divergence is 
known or require to find the parent scalar function whose gradient is known. 
But the task is not very easy, especially in case of finding vector potential 
whose curl is known. Here, `inverse curl', `inverse divergence' and 
`inverse gradient' operators are defined to solve those problems easily. 
\vspace{.25cm}}}

\section{Introduction}
In physical science, usefulness of the concepts of vector and scalar 
potentials is indisputable. They can be used to replace more 
conventional concepts of magnetic and electric fields. In fact, in modern era 
of science, these concepts of potentials became so useful and popular that 
their presence can be seen in virtually all disciplines related somehow to 
`electromagnetism'! The operator which relates those potentials to the 
conventional fields is the so-called vector `del' operator ($\vec{\nabla}$). 
This operator plays a very important role in a wide range of physics besides 
being indispensable in electromagnetism. As for example, it is now very common 
to see the operator in fluid dynamics, quantum mechanics, statistical 
mechanics, classical mechanics, theory of relativity,  thermodynamics, etc.
There are three basic operations, namely, curl, divergence and gradient, which 
can be performed by the operator to relate different quantities of importance.  
For the ease of discussion, let us divide the functions in two categories, 
(a) potential functions (this can be vector or scalar) and (b) field functions 
(again can be vector or scalar). Above mentioned operations are performed on 
potential functions to obtain corresponding field functions. While these 
operations are easy to perform, a back operation to get a potential function 
from a given field function is not always trivial. Since in some cases, it is 
convenient to replace one type of functions with others, one should be able to 
interchange these two types of functions comfortably. Unfortunately, a 
potential function is not unique for a given field function and there is no 
general and unique procedure to obtain a potential function from a given field 
function. In fact, due to this non-uniqueness, one may wonder whether it is at 
all possible to find any general procedure for this purpose. It is exactly the 
concern addressed in this article. Here it is shown that, it is possible to 
find some general procedures in the form of inverse vector operators, which can 
be applied easily to the field functions to obtain the corresponding potential 
functions.

\noindent All the inverse operators are defined here in the orthogonal 
curvilinear co-ordinate system \cite{arfken}. Let $u_1$, $u_2$, $u_3$ be the 
three mutually perpendicular families of co-ordinate surfaces and $\hat{e}_1$, 
$\hat{e}_2$, $\hat{e}_3$ as unit vectors normal to respective surfaces.
\section{Inverse Curl Operator}
At first we mention some symbols, used later.
\renewcommand{\labelenumi}{\alph{enumi}.}
\begin{enumerate}
\item $(\vec{\nabla}\times)^{-1}$ indicates inverse curl operator.\\
\item $(\int^{+}+\int^{-})(\partial u_{1})$ does integration w.r.t. $u_1$, 
while other two variables of the integrand are to be treated fixed. If the 
integrand is the 3rd component (along $\hat{e}_3$) of a vector, then 
$\int^{+}(\partial u_1)$ acts only on the part of the integrand having the 
variable $u_3$. Similarly $\int^{-}(\partial u_1)$ acts only on the part of 
the integrand does not contain the variable $u_3$. For example, let 
$\phi(u_1,u_2,u_3)=\Psi_1(u_1,u_2)+\Psi_2(u_1,u_2,u_3)$ be the component of a 
vector along $\hat{e}_3$, then, $(p_1\int^{+}+p_2\int^{-}) \phi \partial u_{1}
=p_1\int^{+}\Psi_2\partial u_1 + p_2\int^{-}\Psi_1\partial u_1$, with $p_1$ 
and $p_2$ being any two numbers ($u_2, u_3$ are to be treated as constants).
\end{enumerate}
\noindent Now take two vectors $\vec{A}=A_1\hat{e}_1+A_2\hat{e}_2+A_3
\hat{e}_3$ and $\vec{B}=B_1\hat{e}_1+B_2\hat{e}_2+B_3\hat{e}_3$, such that,
\begin{eqnarray}
\vec{B}=\vec{\nabla}\times\vec{A}
\label{e1}
\end{eqnarray}
\noindent Using inverse curl operator, this vector potential $\vec{A}$ can be 
expressed as,
\begin{eqnarray}
\vec{A}=(\vec{\nabla}\times)^{-1}\vec{B}
\label{e2}
\end{eqnarray}
\noindent From eqn (\ref{e1}) it is obvious that,
\begin{eqnarray}
& &\vec{\nabla}\cdot\vec{B}=\vec{\nabla}\cdot(\vec{\nabla}\times\vec{A}) = 0
\nonumber \\
&\Rightarrow & \frac{1}{h_1h_2h_3}[\frac{\partial}{\partial u_1}(h_2h_3B_1)+
\frac{\partial}{\partial u_2}(h_3h_1B_2)+\frac{\partial}{\partial u_3}
(h_1h_2B_3)]=0
\nonumber \\
&\Rightarrow & \frac{\partial}{\partial u_1}[c_1(u_1^{+})+c_1(u_1^{-})]+
\frac{\partial}{\partial u_2}[c_2(u_2^{+})+c_2(u_2^{-})]+
\frac{\partial}{\partial u_3}[c_3(u_3^{+})+c_3(u_3^{-})]=0
\nonumber \\
&\Rightarrow & \frac{\partial}{\partial u_1}c_1(u_1^{+})+
\frac{\partial}{\partial u_2}c_2(u_2^{+})+
\frac{\partial}{\partial u_3}c_3(u_3^{+})=0
\label{e3}
\end{eqnarray}
\noindent where $c_1(u_1^{+})$ and $c_1(u_1^{-})$ are two parts of the term 
$h_2h_3B_1$, containing and not containing $u_1$ respectively, such that, 
$h_2h_3B_1=c_1(u_1^{+})+c_1(u_1^{-})$. Other terms can be defined similarly.
\noindent From eqn (\ref{e1}) we get,
\begin{eqnarray}
\vec{B}=\vec{\nabla}\times\vec{A}=
\begin{array}{c|ccc|}
 & h_1\hat{e}_1 & h_2\hat{e}_2 & h_3\hat{e}_3 \\
 \frac{1}{h_1h_2h_3} & \frac{\partial}{\partial u_1} & 
\frac{\partial}{\partial u_2} & \frac{\partial}{\partial u_3} \\
 & h_1A_1 & h_2A_2 & h_3A_3 
\end{array}
\label{e4}
\end{eqnarray}

\begin{eqnarray}
 \Rightarrow B_1\hat{e}_1+B_2\hat{e}_2+B_3\hat{e}_3 & = & 
\frac{\hat{e}_1}{h_2h_3}[\frac{\partial}{\partial u_2}(h_3A_3)-\frac{\partial}
{\partial u_3}(h_2A_2)]\nonumber\\
 & + & \frac{\hat{e}_2}{h_3h_1}[\frac{\partial}{\partial u_3}(h_1A_1)-
\frac{\partial}{\partial u_1}(h_3A_3)]\nonumber\\
 & + & \frac{\hat{e}_3}{h_1h_2}[\frac{\partial}{\partial u_1}(h_2A_2)-
\frac{\partial}{\partial u_2}(h_1A_1)]
\label{e5}
\end{eqnarray}
\noindent We already have defined components of $\vec{B}$ in terms of $c$'s. 
Now, we have to choose $h_1A_1$, $h_2A_2$ and $h_3A_3$ in terms of $c$'s in 
such a way that, {\it r.h.s.} of the eqn (\ref{e5}) becomes same as 
{\it l.h.s.}
\vspace{.6cm}\\
\noindent Let (by inspection),

$h_1A_1=k_1\int c_2(u_2^{+})\partial u_3+k_2\int c_3(u_3^{+})\partial u_2+
k_3\int c_2(u_2^{-})\partial u_3+k_4\int c_3(u_3^{-})\partial u_2$

\noindent Similarly,

$h_2A_2=k_5\int c_3(u_3^{+})\partial u_1+k_6\int c_1(u_1^{+})\partial u_3+
k_7\int c_3(u_3^{-})\partial u_1+k_8\int c_1(u_1^{-})\partial u_3$

\noindent and,
$h_3A_3=k_9\int c_1(u_1^{+})\partial u_2+k_{10}\int c_2(u_2^{+})\partial u_1+
k_{11}\int c_1(u_1^{-})\partial u_2+k_{12}\int c_2(u_2^{-})\partial u_1$

\noindent where, $k_1, k_2, \cdots, k_{12}$ are some constants to be 
determined.

\noindent (Note that, $c_1$ terms are not included for $h_1A_1$, since, 
$h_1A_1$ does not appear in the expression of 1st component $B_1$ in 
eqn (\ref{e5}). Also note that, terms in the expression of $h_1A_1$ 
are in the integral form {\it w.r.t.} variables $u_2$ and $u_3$ but not $u_1$, 
as in eqn (\ref{e5}), $h_1A_1$ is never differentiated {\it w.r.t.} $u_1$.)
\vspace{.6cm}\\
\noindent Now, see the coefficient of $\hat{e}_3$ (expression of $B_3$) on the 
{\it r.h.s.} of eqn (\ref{e5}),
~\\

$\frac{1}{h_1h_2}[\frac{\partial}{\partial u_1}(h_2A_2)-
\frac{\partial}{\partial u_2}(h_1A_1)]$
\vspace{.5cm}\\
$=\frac{1}{h_1h_2}[\frac{\partial}{\partial u_1}\{k_5\int c_3(u_3^{+})
\partial u_1+k_6\int c_1(u_1^{+})\partial u_3+k_7\int c_3(u_3^{-})
\partial u_1+k_8\int c_1(u_1^{-})\partial u_3\}\\\hspace*{.9cm}-\frac{\partial}
{\partial u_2}\{k_1\int c_2(u_2^{+})\partial u_3+k_2\int c_3(u_3^{+})
\partial u_2+k_3\int c_2(u_2^{-})\partial u_3+k_4\int c_3(u_3^{-})
\partial u_2\}]$
\vspace{.5cm}\\
$=\frac{1}{h_1h_2}[k_5c_3(u_3^{+})+k_6\int\frac{\partial}{\partial u_1}
c_1(u_1^{+})\partial u_3+k_7c_3(u_3^{-})+k_8\cdot0-k_1\int\frac{\partial}
{\partial u_2}c_2(u_2^{+})\partial u_3 \\ 
\hspace*{.9cm}-k_2c_3(u_3^{+})-k_3\cdot0-k_4c_3(u_3^{-})]$
\vspace{.5cm}\\
$=\frac{1}{h_1h_2}[2k_5c_3(u_3^{+})-k_1\int\{\frac{\partial}{\partial u_1} c_1
(u_1^{+})+\frac{\partial}{\partial u_2}c_2(u_2^{+})\}\partial u_3+2k_7c_3
(u_3^{-})]$\\
\hspace*{8.7cm}[Taking $k_5=-k_2$, $k_7=-k_4$ and $k_6=-k_1$]
\vspace{.3cm}\\
$=\frac{1}{h_1h_2}[2k_5c_3(u_3^{+})-k_1\int\{-\frac{\partial}
{\partial u_3}c_3(u_3^{+})\}\partial u_3+2k_7c_3(u_3^{-})]$ 
\hspace*{3.7cm} [Using eqn (\ref{e3})]
\vspace{.5cm}\\
$=\frac{1}{h_1h_2}[2k_5c_3(u_3^{+})+k_1c_3(u_3^{+})+2k_7c_3(u_3^{-})]$
\vspace{.5cm}\\
$=\frac{1}{h_1h_2}[3k_5c_3(u_3^{+})+2k_7c_3(u_3^{-})]$ \hspace*{7.7cm} [Taking 
$k_1=k_5$]
\vspace{.5cm}\\
$=\frac{1}{h_1h_2}[3\cdot\frac{1}{3}c_3(u_3^{+})+2\cdot\frac{1}{2}c_3
(u_3^{-})]$ 
\hspace*{4.1cm}[Considering $k_5=1/3$ and $k_7=1/2$]
\vspace{.5cm}\\
$=\frac{1}{h_1h_2}[c_3(u_3^{+})+c_3(u_3^{-})]$
\vspace{.5cm}\\
$=\frac{1}{h_1h_2}(h_1h_2B_3)$
\vspace{.5cm}\\
$=B_3$
\vspace{.5cm}\\
\noindent Similarly we can show that, if,

\noindent $k_1=k_5=k_9=-k_2=-k_6=-k_{10}=1/3$ \\and,
\noindent $k_3=k_7=k_{11}=-k_4=-k_8=-k_{12}=1/2$, then,

\noindent $\frac{1}{h_2h_3}[\frac{\partial}{\partial u_2}(h_3A_3)-
\frac{\partial}{\partial u_3}(h_2A_2)]=B_1$ and 
$\frac{1}{h_1h_3}[\frac{\partial}{\partial u_3}(h_1A_1)-
\frac{\partial}{\partial u_1}(h_3A_3)]=B_2$
\vspace{.2cm}\\

\noindent So, using these values of $k_1, k_2, \cdots, k_{12}$, we get,

\noindent $A_1=\frac{1}{3h_1}[\int c_2(u_2^{+})\partial u_3-\int 
c_3(u_3^{+})\partial u_2]+\frac{1}{2h_1}[\int c_2(u_2^{-})\partial 
u_3-\int c_3(u_3^{-})\partial u_2]$
\vspace{.1cm}\\

\noindent $A_2=\frac{1}{3h_2}[\int c_3(u_3^{+})\partial u_1-\int 
c_1(u_1^{+})\partial u_3]+\frac{1}{2h_2}[\int c_3(u_3^{-})\partial 
u_1-\int c_1(u_1^{-})\partial u_3]$
\vspace{.1cm}\\

\noindent $A_3=\frac{1}{3h_3}[\int c_1(u_1^{+})\partial u_2-\int 
c_2(u_2^{+})\partial u_1]+\frac{1}{2h_3}[\int c_1(u_1^{-})\partial 
u_2-\int c_2(u_2^{-})\partial u_1]$
\vspace{.1cm}\\

\noindent Now from eqn (\ref{e2}), we get,\\
$(\vec{\nabla}\times)^{-1}\vec{B}=\vec{A}=A_1\hat{e}_1+A_2\hat{e}_2+
A_3\hat{e}_3$
\vspace{.5cm}\\
$=\frac{\hat{e}_1}{3h_1}[\int c_2(u_2^{+})\partial u_3-\int c_3(u_3^{+})
\partial u_2]+\frac{\hat{e}_1}{2h_1}[\int c_2(u_2^{-})\partial u_3-\int 
c_3(u_3^{-})\partial u_2]$ \vspace{.3cm}\\\hspace*{.15cm}
$+\frac{\hat{e}_2}{3h_2}[\int c_3(u_3^{+})\partial u_1-\int c_1(u_1^{+})
\partial u_3]+\frac{\hat{e}_2}{2h_2}[\int c_3(u_3^{-})\partial u_1-\int 
c_1(u_1^{-})\partial u_3]$ \vspace{.3cm}\\\hspace*{.15cm}
$+\frac{\hat{e}_3}{3h_3}[\int c_1(u_1^{+})\partial u_2-\int c_2(u_2^{+})
\partial u_1]+\frac{\hat{e}_3}{2h_3}[\int c_1(u_1^{-})\partial u_2-\int 
c_2(u_2^{-})\partial u_1]$
\vspace{.5cm}\\
$=-\frac{\hat{e}_1}{3h_1} \begin{array}{|cc|} \int(\partial u_2) & 
\int(\partial u_3) \\ c_2(u_2^{+}) & c_3(u_3^{+}) \end{array} + 
\frac{\hat{e}_2}{3h_2} \begin{array}{|cc|} \int(\partial u_1) &
\int(\partial u_3) \\ c_1(u_1^{+}) & c_3(u_3^{+}) \end{array} -
\frac{\hat{e}_3}{3h_3} \begin{array}{|cc|} \int(\partial u_1) &
\int(\partial u_2) \\ c_1(u_1^{+}) & c_2(u_2^{+}) \end{array}
\vspace{.5cm}\\\hspace*{.25cm}
-\frac{\hat{e}_1}{2h_1} \begin{array}{|cc|} \int(\partial u_2) &
\int(\partial u_3) \\ c_2(u_2^{-}) & c_3(u_3^{-}) \end{array} +
\frac{\hat{e}_2}{2h_2} \begin{array}{|cc|} \int(\partial u_1) &
\int(\partial u_3) \\ c_1(u_1^{-}) & c_3(u_3^{-}) \end{array} -
\frac{\hat{e}_3}{2h_3} \begin{array}{|cc|} \int(\partial u_1) &
\int(\partial u_2) \\ c_1(u_1^{-}) & c_2(u_2^{-}) \end{array}$
\vspace{.5cm}\\
$=-\frac{1}{3} \begin{array}{|ccc|} \frac{\hat{e}_1}{h_1} & 
\frac{\hat{e}_2}{h_2} & \frac{\hat{e}_3}{h_3}\\ & & \\ 
\int(\partial u_1) & 
\int(\partial u_2) & \int(\partial u_3) \\ & & \\ 
c_1(u_1^{+}) & c_2(u_2^{+}) & c_3(u_3^{+}) \end{array}
-\frac{1}{2} \begin{array}{|ccc|} \frac{\hat{e}_1}{h_1} &
\frac{\hat{e}_2}{h_2} & \frac{\hat{e}_3}{h_3}\\ & & \\
\int(\partial u_1) &
\int(\partial u_2) & \int(\partial u_3) \\ & & \\
c_1(u_1^{-}) & c_2(u_2^{-}) & c_3(u_3^{-}) \end{array}$
\vspace{.5cm}\\
$=- \begin{array}{|ccc|} \frac{\hat{e}_1}{h_1} &
\frac{\hat{e}_2}{h_2} & \frac{\hat{e}_3}{h_3}\\ & & \\
\int(\partial u_1) &
\int(\partial u_2) & \int(\partial u_3) \\ & & \\
\frac{1}{3}c_1(u_1^{+})+\frac{1}{2}c_1(u_1^{-}) & 
\frac{1}{3}c_2(u_2^{+})+\frac{1}{2}c_2(u_2^{-}) & 
\frac{1}{3}c_3(u_3^{+})+\frac{1}{2}c_3(u_3^{-}) \end{array} $
\vspace{.6cm}\\
$=-\begin{array}{|ccc|} \frac{\hat{e}_1}{h_1} &
\frac{\hat{e}_2}{h_2} & \frac{\hat{e}_3}{h_3}\\ & & \\
(\frac{1}{3}\int^{+}+\frac{1}{2}\int^{-})(\partial u_1) &
(\frac{1}{3}\int^{+}+\frac{1}{2}\int^{-})(\partial u_2) &
(\frac{1}{3}\int^{+}+\frac{1}{2}\int^{-})(\partial u_3) \\ & & \\
h_2h_3B_1 & h_3h_1B_2 & h_1h_2B_3 \end{array} $ 
\vspace{.4cm}\\
\noindent This is the desired expression for our inverse curl operator.\\
Since, $\vec{\nabla}\times(\vec{\nabla}\phi)=0$, so, it should be noted that, 
for a given solenoidal vector, corresponding vector potential is not unique, 
which can be in general expressed as, -\vspace{.2cm}\\
$\vec{A}=(\vec{\nabla}\times)^{-1}\vec{B}+\vec{\nabla}\phi$
\vspace{.2cm}\\
where $\phi$ is an arbitrary scalar function.
\subsection{An Example in Cartesian System}
One example in Cartesian system would clarify the procedure of getting vector 
potential for a solenoidal vector. In this system, $\hat{e}_1$, $\hat{e}_2$, 
$\hat{e}_3$, $u_1$, $u_2$ and $u_3$ are respectively replaced by $\hat{i}$, 
$\hat{j}$, $\hat{k}$, $x$, $y$, and $z$. Here $h_1=h_2=h_3=1$.

\noindent Now take a solenoidal vector $\vec{B}=(xyz+y^{2})\hat{i}+(xz+y)
\hat{j}-z(1+yz/2)\hat{k}$. (note $\vec{\nabla}\cdot\vec{B}=0$) 

\noindent So, vector potentials of $\hat{B}$ are,-\vspace{.2cm}\\ 
$\vec{A}=(\vec{\nabla}\times)^{-1}\vec{B}+\vec{\nabla}\phi$ \vspace{.4cm}\\
$=-\begin{array}{|ccc|} \hat{i} & \hat{j} & \hat{k} \\ & & \\
(\frac{1}{3}\int^{+}+\frac{1}{2}\int^{-})(\partial x) &
(\frac{1}{3}\int^{+}+\frac{1}{2}\int^{-})(\partial y) & 
(\frac{1}{3}\int^{+}+\frac{1}{2}\int^{-})(\partial z) \\ & & \\
xyz+y^{2} & xz+y & -z-yz^{2}/2 \end{array}+\vec{\nabla}\phi$ \vspace{.4cm}\\
$=-\hat{i}[\frac{1}{3}\int(-z-yz^{2}/2)\partial y+\frac{1}{2}\int0\partial y -
\frac{1}{3}\int y\partial z -\frac{1}{2}\int xz\partial z] \vspace{.4cm}\\
\hspace*{.25cm}-\hat{j}[\frac{1}{3}\int xyz\partial z +\frac{1}{2}\int y^{2} 
\partial z -\frac{1}{3}\int(-z-yz^{2}/2)\partial x -\frac{1}{2}\int0\partial x]
 \vspace{.4cm}\\\hspace*{.25cm} -\hat{k}[\frac{1}{3}\int y\partial x + 
\frac{1}{2}\int xz \partial x -\frac{1}{3}\int xyz \partial y -\frac{1}{2} 
\int y^{2} \partial y] + \vec{\nabla}\phi$ \vspace{.4cm}\\
$=\hat{i}[(zy+z^{2}y^{2}/4)/3+(yz/3+xz^{2}/4)]-\hat{j}[(zx+yxz^{2}/2)/3+
(xyz^{2}/6+y^{2}z/2)]\\\hspace*{.25cm}+\hat{k}[-(yx/3+x^{2}z/4)+xy^{2}z/6+
y^{3}/6]+\vec{\nabla}\phi$ 
\subsubsection{ Verification of the Result}
$\vec{\nabla}\times\vec{A}=\begin{array}{|ccc|} 
\hat{i} & \hat{j} & \hat{k} \\ & & \\
\frac{\partial}{\partial x} & \frac{\partial}{\partial y} & 
\frac{\partial}{\partial z} \\ & & \\
(zy+z^{2}y^{2}/4)/3 & -(zx+xyz^{2}/2)/3 & -(yx/3+x^{2}z/4)\\
+(yz/3+xz^{2}/4) & -(xyz^{2}/6+y^{2}z/2) & +(xy^{2}z/6+y^{3}/6) \end{array} 
+\vec{\nabla}\times(\vec{\nabla}\phi)$ \vspace{.4cm}\\\hspace*{1.0cm}
$=\hat{i}(xyz+y^{2})+\hat{j}(xz+y)-\hat{k}z(1+yz/2)+0$
\vspace{.4cm}\\\hspace*{1.0cm}
$=\vec{B}$
\section{Inverse Divergence Operator}
Here we indicate this operator by $(\vec{\nabla}\cdot)^{-1}$. Now if $\vec{A}$ 
and $\phi$ be a vector and a scalar function respectively, such that $\phi=
\vec{\nabla}\cdot\vec{A}=\frac{1}{h_1h_2h_3}[\frac{\partial}{\partial u_1}
(h_2h_3A_1)+\frac{\partial}{\partial u_2}(h_1h_3A_2)+\frac{\partial}
{\partial u_3}(h_1h_2A_3)]$, then $\vec{A}$ can be expressed as 
$\vec{A}=(\vec{\nabla}\cdot)^{-1}\phi$.
~\\

\noindent We can now define the operator as (by inspection),

\noindent $(\vec{\nabla}\cdot)^{-1}=k_1\frac{\hat{e}_1}{h_2h_3}\int h_1h_2h_3
(\partial u_1)+ k_2\frac{\hat{e}_2}{h_3h_1}\int h_1h_2h_3(\partial u_2)+
k_3\frac{\hat{e}_3}{h_1h_2}\int h_1h_2h_3
(\partial u_3)$

\noindent where, $k_1+k_2+k_3=1$.
\subsection{Verification}
For a given $\phi$,
$\vec{A}=(\vec{\nabla}\cdot)^{-1}\phi$\vspace{.3cm}\\\hspace*{.2cm}
$=k_1\frac{\hat{e}_1}{h_2h_3}\int h_1h_2h_3\phi
(\partial u_1)+
k_2\frac{\hat{e}_2}{h_3h_1}\int h_1h_2h_3\phi
(\partial u_2)+
k_3\frac{\hat{e}_3}{h_1h_2}\int h_1h_2h_3\phi
(\partial u_3)$\vspace{.3cm}\\
Now note that, divergence of the obtained vector gives the scalar function 
$\phi$,\vspace{.3cm}\\
$\vec{\nabla}\cdot\vec{A}=\frac{1}{h_1h_2h_3}[k_1h_1h_2h_3\phi+
k_2h_1h_2h_3\phi+k_3h_1h_2h_3\phi]\vspace{.3cm}\\\hspace*{.2cm}
=(k_1+k_2+k_3)\phi=\phi$\vspace{.3cm}\\
Since, $\vec{\nabla}\cdot(\vec{\nabla}\times\vec{B})=0$, so, it should be 
noted that for a given scalar function, corresponding potential function is not 
unique, which can be in general expressed as,\vspace{.3cm}\\
$\vec{A}=(\vec{\nabla}\cdot)^{-1}\phi+\vec{\nabla}\times\vec{B}$\vspace{.2cm}\\
where $\vec{B}$ is any vector.
\section{Inverse Gradient Operator}
We indicate this operator by $(\vec{\nabla})^{-1}$.

\noindent Let $\vec{A}=A_1\hat{e}_1+A_2\hat{e}_2+A_3\hat{e}_3$ and 
$\phi(u_1,u_2,u_3)$ be a vector and a scalar function, such that 
$\vec{A}=\vec{\nabla}\phi$.

\noindent Since, the vector field represented by $\vec{A}$ is conservative, 
thus, line integral of $\vec{A}$ is path independent. Choose (a,b,c) as 
initial point, such that $\phi(a,b,c)$ exists.

\noindent Now $\phi(u_1,u_2,u_3)=\begin{array}{c} (u_1,u_2,u_3)\\\int 
d\phi\\(a,b,c) \end{array}+c_0$ \hspace*{1cm} [$c_0$ is integration constant]
\vspace{.3cm}\\\hspace*{3.4cm} 
$=\begin{array}{l} (u_1,u_2,u_3)\\\int(\vec{\nabla}\phi)\cdot\vec{dr}\\
(a,b,c)\end{array} +c_0$\vspace{.3cm}\\\hspace*{3.4cm}
$=\begin{array}{l} (u_1,u_2,u_3)\\\int\vec{A}\cdot\vec{dr}\\
(a,b,c)\end{array} +c_0$\vspace{.3cm}\\\hspace*{3.4cm}
$=\begin{array}{l}(a,b,u_3)\\\int A_3h_3du_3\\(a,b,c) \end{array}+
\begin{array}{l}(a,u_2,u_3)\\\int A_2h_2du_2\\(a,b,u_3) \end{array}+
\begin{array}{l}(u_1,u_2,u_3)\\\int A_1h_1du_1\\(a,u_2,u_3) \end{array}+c_0$
\vspace{.3cm}\\
So, \\
$\phi=(\vec{\nabla})^{-1}\vec{A}\vspace{.3cm}\\\hspace*{.3cm}
=\begin{array}{c} u_1\\\int\\a\end{array}A_1h_1\partial u_1 +
\begin{array}{c} u_2\\\int\\b\end{array}A_2h_2\partial u_2 +
\begin{array}{c} u_3\\\int\\c\end{array}A_3h_3\partial u_3+c_0$\\
\hspace*{3.4cm}$u_1=a$ \hspace*{1.3cm}$\begin{array}{c}u_1=a\\u_2=b 
\end{array}$\vspace{.2cm}\\
Here in the first integration $u_1$ is variable but other two ($u_2, u_3$) are
 fixed, and so on for other two integrations.

\pagebreak
{\noindent
\vspace{.2cm}\\
{\bf \it Note:} The original work was done in 2002 and published in 2006 in 
Science and Culture \cite{sahoo}. This write-up is almost same as 
published one. 
}

\end{document}